\documentclass[aps,pra,superscriptaddress,nofootinbib,notitlepage,showpacs,floatfix,twocolumn]{revtex4-1}

\usepackage[utf8]{inputenc}
\usepackage[english]{babel}
\usepackage{mathtools}
\usepackage{mathptmx}

\begin{document}
\title{Optimal Circuit-Level Decoding for Surface Codes}
\author{Bettina Heim}
\affiliation{Theoretische Physik, ETH Zurich, 8093 Zurich, Switzerland}
\author{Krysta M. Svore}
\affiliation{Station Q Quantum Architectures and Computation Group, Microsoft Research, Redmond, WA 98052, USA}
\author{Matthew B. Hastings}
\affiliation{Station Q, Microsoft Research, Santa Barbara, CA 93106-6105, USA}
\affiliation{Station Q Quantum Architectures and Computation Group, Microsoft Research, Redmond, WA 98052, USA}
\begin{abstract}
Surface codes exploit topological protection to increase error resilience in quantum computing devices and can in principle be implemented in existing hardware. They are one of the most promising candidates for active error correction, not least due to a polynomial-time decoding algorithm which admits one of the highest predicted error thresholds.
We consider the dependence of this threshold on underlying assumptions including different noise models, and analyze the performance of a minimum weight perfect matching (MWPM) decoding compared to a mathematically optimal maximum likelihood (ML) decoding. Our ML algorithm tracks the success probabilities for all possible corrections over time and accounts for individual gate failure probabilities and error propagation due to the syndrome measurement circuit. 
We present the very first evidence for the true error threshold of an optimal circuit level decoder, allowing us to draw conclusions about what kind of improvements are possible over standard MWPM. 
\end{abstract}
\maketitle

Topological stabilizer codes, such as toric codes \cite{Kitaev2003} and related surface codes, promise fault-tolerant quantum computation by encoding logical information into global degrees of freedom that are insensitive to local perturbations. 
While local interactions result in the formation and movement of quasiparticles across the system \cite{Brell2016, Kitaev2003, nayak2008}, only their accidental braiding corrupts stored information \cite{Kitaev2003, Lloyd2003}.
Based on the syndromes obtained from measurements of local stabilizer operators, a classical decoding algorithm tries to predict the correct logical state by tracking their most likely locations.
As this corresponds to an NP-hard optimization problem \cite{landahl2011, poulin2006}, profiting from the computation speed of a quantum computer necessitates the use of a clever approximate decoder.

For surface and toric codes, a polynomial-time algorithm for syndrome evaluation exists based on minimum weight perfect matching (MWPM) \cite{Dennis2002, cook1999, efficientMWPMimplementation}.
It promises one of the highest error thresholds~\cite{wang2011surface,WFcolor} for reliable computation: roughly $1\%$ under certain noise types.
Even though the logical error rate decays exponentially with code distance as long as the noise is 
below threshold, in practice one may need an enormous number of physical qubits to 
	sufficiently protect information close to threshold. The ability to decode efficiently and robustly for any type of noise can therefore have a critical effect on logical qubit and gate reliability.
	
Until now, it is neither understood how much a better decoding can increase predicted threshold values, nor how correlated noise impacts code performance.
While the thresholds for perfect syndrome extraction or for uncorrelated syndrome bit errors can be determined by finding a phase transition in a random bond Ising and random bond gauge model, respectively \cite{Dennis2002}, 
	these simplified models do not incorporate how faulty measurements and excitations are related in an experimental setting, where each gate in the syndrome extraction circuit can cause, propagate and modify errors. Such correlations need 
	to be accounted for not only when estimating the code performance in hardware, but also in the decoder.

This letter highlights what kind of improvements over standard MWPM are conceivable using knowledge about the occurring circuit noise. We present the very first simulations of surface codes using a mathematically optimal maximum likelihood decoder based on a circuit-level noise model (MLCLN). 
	Going beyond the phenomenological model in Ref.~\onlinecite{ColorCodesML, Nickerson2016}, our decoder fully accounts for individual gate failure rates and error propagation. 
	Our simulations provide insight that can guide the development of future enhanced decoding algorithms by identifying the noise events and correlations that are most vital to capture for a more efficient decoding.

We estimate the surface code threshold under circuit noise for MWPM and MLCLN and find that there is a non-negligible difference between the optimal threshold and the one using MWPM, with an even larger relative advantage for MLCLN when considering strongly correlated noise. 
	For both decodings the error threshold depends significantly on the noise model. For independent bit- and phase-flip errors on each gate, it is as low as 0.35\% even with optimal decoding, whereas for the noise model in Ref.~\onlinecite{WangFowler} a threshold as high as 1.8\% may be achievable.

{\it Surface Codes---}
We briefly explain the key ideas of stabilizer codes and syndrome extraction before introducing maximum likelihood and MWPM decoding. We refer the reader to the supplemental material for detail and to Refs.~\onlinecite{SurfaceCodesFowler, Fowler2012} for an introduction to surface codes.

The stabilizer formalism describes a quantum system by a set of operators under which the state vector is invariant. For surface codes, these so-called stabilizers generate a group $\mathcal{S}$ of commuting Pauli operators, and their simultaneous $+1$ eigenspace is isomorphic to the Hilbert space of the encoded \emph{logical qubit(s)}, called the code space. 
In this letter, we encode one logical qubit into highly entangled states on a grid of $n$ physical qubits, called \emph{data qubits}. The stabilizer generators are parity checks of the form $\prod_{i} X_{i}$ and $\prod_{i} Z_{i}$ between square cells of four neighboring data qubits. Projective measurements onto their eigenspaces allow to detect local excitations. Each measurement is implemented via a circuit consisting of four $CNOT$ gates between the involved data qubits and an additional ancilla qubit, followed by a measurement of the latter. 
	The entire syndrome extraction circuit that determines the \emph{measured syndrome} 
	requires $n_a$ ancilla qubits, with $|{\mathcal S}|=2^{n_a}$.
	As each gate can be faulty, the extracted parity values are not necessarily correct. 
	
We associate the data qubit excitation pattern after syndrome extraction with the \emph{ideal syndrome} defined by the $n_a$ eigenvalues of the stabilizer generators. It determines the set of Pauli operators that map the system back to the code space, though not necessarily back onto the correct logical state. 
	After projection, any excitation of the extended system containing data and ancilla qubits can therefore be characterized by three variables: an ideal syndrome reflecting the parity of the data qubit excitations,
		an imperfect measured syndrome corresponding to the state of the ancilla qubits, 
		and a logical operator accounting for undetectable global excitations, giving $4\cdot 2^{2n_a}=4|\mathcal{S}|^2$ possibilities.
		We will call the set of errors identified by a combination of these three values an EC-coset. 

{\it Decoding---}
Starting from an error-free system and given a noise model detailing the probabilities for all $4^{n+n_a}$  possible Pauli errors to occur during each step of the syndrome extraction circuit, we can calculate the probability of an error after its execution to belong to any one of the EC-cosets. This 'one-step probability vector' defines a transition matrix between EC-cosets, as it defines the transition probability between any two cosets differing by a certain bit pattern. Given the measured syndrome, the propagation is described in its entirety by a ${4|\mathcal{S}| \times 4|\mathcal{S}|}$ matrix (see supplemental material). 
	At the end of a quantum algorithm, the measurement of a logical operator in combination with the coset probabilities extracts the computed value. Additional information about its correctness can be gained from a final measurement of all data qubits; indeed, when we determine the logical error rate we check whether the decoder would predict the correct logical state given that additional information.
	
Such a MLCLN decoder becomes intractable with growing code distance, and an approximate scheme for decoding is needed in practice. 
Avoiding the problem of tracking an exponentially scaling number of probabilities, MWPM instead searches
for the minimum in an energy landscape defined by the history of syndromes; defects are treated as vertices of a graph on which MWPM finds a perfect matching that minimizes the set weights \cite{Dennis2002}. 
	Correspondingly, the means to identify errors resulting in a whole collection of defects are limited. 
	Setting the edge weights to that of the minimal error chain linking a pair of syndrome bit flips results in a phenomenological decoding that is independent on noise type and strength.
Even though more elaborate weights \cite{wang2011surface,fowler2011accurate} promise a more reliable decoding, correlations between multiple error chains remain largely neglected, as accounting for them results in an increased classical complexity. 
This leads to two major questions we will address in this letter: 
	How important is it to account for noise correlations that arise from the syndrome extraction circuit? 
	What type of noise leads to poor decoding using standard MWPM and how much can the use of knowledge about the  noise occurring improve decoding?

\begin{figure}
	\hspace*{-0.22cm}\includegraphics[width=1.032\columnwidth]{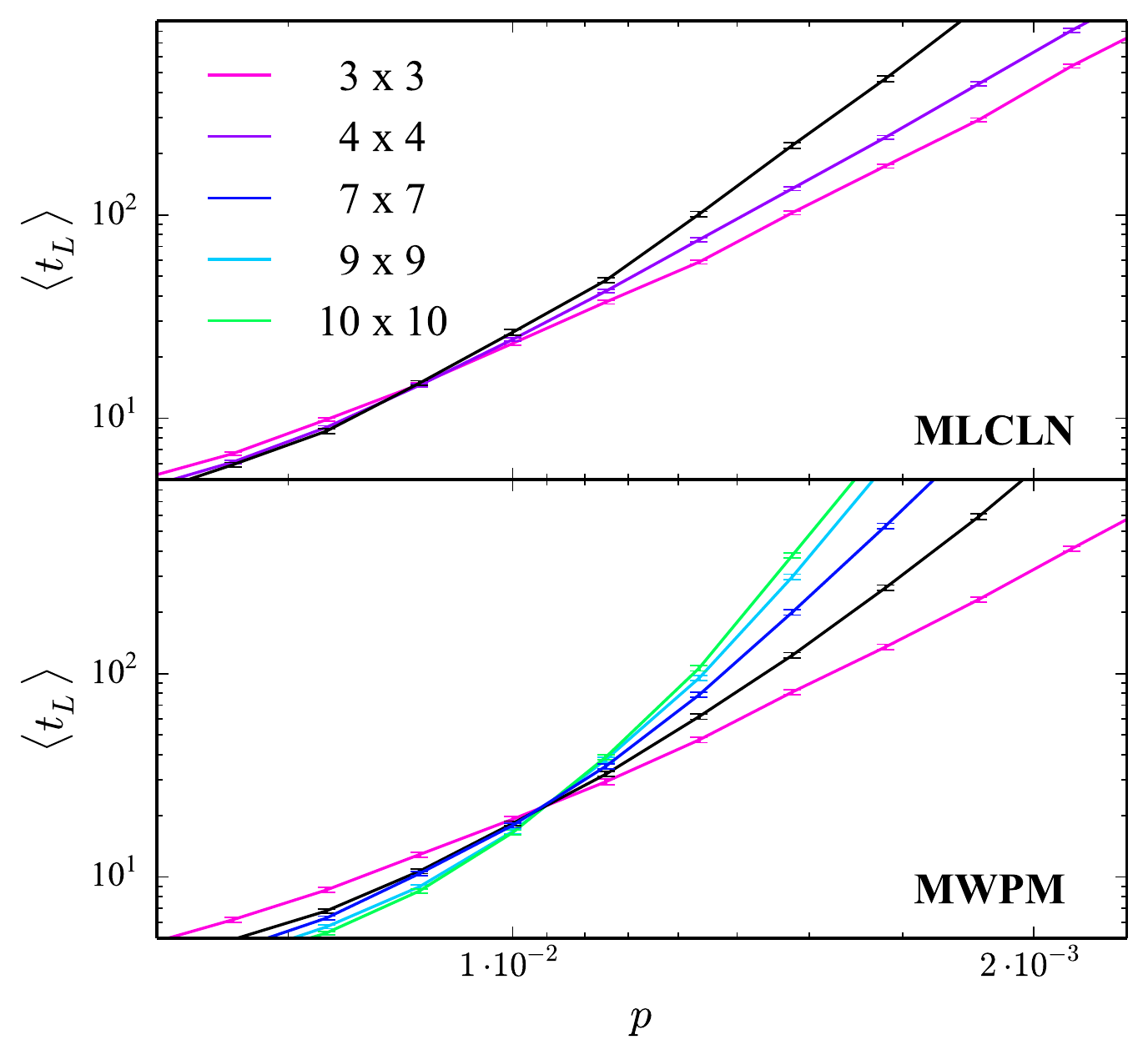}	
	\vspace*{-0.7cm}\caption{Error threshold curves for various distance surface codes using MWPM and MLCLN decoders, with correlated 2-qubit errors on $CNOT$ gates and depolarizing noise on single-qubit gates, for a 6-step syndrome measurement circuit that minimizes error propagation (SN-circuit).}
	\label{correlated_noise}
	\vspace{-0.5cm}
\end{figure}

{\it Implementation---}
To answer these questions we simulate the time evolution of a quantum memory while continuously performing syndrome measurements.
Errors are generated according to a given noise model and applied after each gate during the syndrome extraction circuit. After each completed cycle the decoder is queried to determine whether the logical state is still correct. The simulation is terminated and repeated as soon as its prediction is incorrect. Note that corrections never actually need to be applied; tracking a so-called ``Pauli frame" \cite{pframe1,pframe2} is enough. In our plots, $p$ denotes the error rate given by the noise model and $\langle t_L\rangle$ denotes the average number of syndrome measurement cycles performed before a logical error occurred. Between 1000 and 2000 simulations are performed for each data point and error bars are obtained by bootstrapping.
A description of the exact decoding algorithms and their implementations are given in the supplemental material, where we also detail the surface code layout and syndrome extraction circuits, and explain numerical optimizations used to render simulations possible.

{\it Error Threshold---}		
The threshold, below which arbitrary length quantum computation is possible, depends on the error correction code, noise type, and decoder.
	In principle it can be estimated by relating the error correction scheme to a classical statistical-mechanical spin model \cite{Dennis2002,katzgraber2013}, where it represents a phase transition to a symmetry-broken phase. 
In practice, numerical tools are crucial for evaluating the threshold under more complex models.
Simulations focusing on surface code thresholds with MWPM under selected noise models have been done in Refs.~\onlinecite{Raussendorf2007threshold, WangFowler, wang2011surface}. Our goal is to determine how it is affected by noise type and circuit details, and how much it can be improved using a better decoder. 

{\it Threshold Dependence on Noise Model---}
To illustrate the extent to which the threshold depends on presumed assumptions, we compare MWPM and MLCLN for two different noise models, one with correlated noise~\cite{WangFowler} using a 6-step circuit and one with independent bit- and phase-flips where the only correlations stem from error propagation through an 8-step circuit. Additional models are in the supplemental material.

We start with the correlated model in Ref.~\onlinecite{WangFowler}. For single-qubit gates, the probability of an $X$, $Y$ or $Z$ error, respectively, is $p/3$, and the probability for each non-trivial two-qubit Pauli error on $CNOT$-gates is $p/15$. The syndrome measurement circuit consists of six steps. First, ancilla qubits belonging to $Z$-type ($X$-type) stabilizers are initialized in a $|0\rangle$ ($|+\rangle$) state. Then four steps of $CNOT$ gates generate the appropriate entanglement between ancilla and data qubits, and in a last step the ancilla qubits are measured with respect to $Z$- or $X$-basis. 

As shown in Fig.~\ref{correlated_noise}, this circuit and noise model lead to a threshold value of around 0.9\% for MWPM. 
We obtain a slightly higher value compared to that in Ref.~\onlinecite{WangFowler} since we weight time-like errors between two subsequent rounds of syndrome measurements only half as much as neighboring spatial errors in order to account for noise propagation caused by the circuit. Other minor differences can be attributed to the fact that we use a slightly different surface code layout which requires a smaller number of qubits. 
We estimate that the threshold value under a mathematically optimal decoder is between 1.5\% and 1.8\%. 
Even though the MLCLN panel in Fig.~\ref{correlated_noise} suggests a crossing near 1.5\%, this value is somewhat below 
the true optimal threshold since we made the approximation of tracking separate marginals of the distribution on $X$ and $Z$ errors due to limited computational resources. 

\begin{figure}
	\hspace*{-0.22cm}\includegraphics[width=1.04\columnwidth]{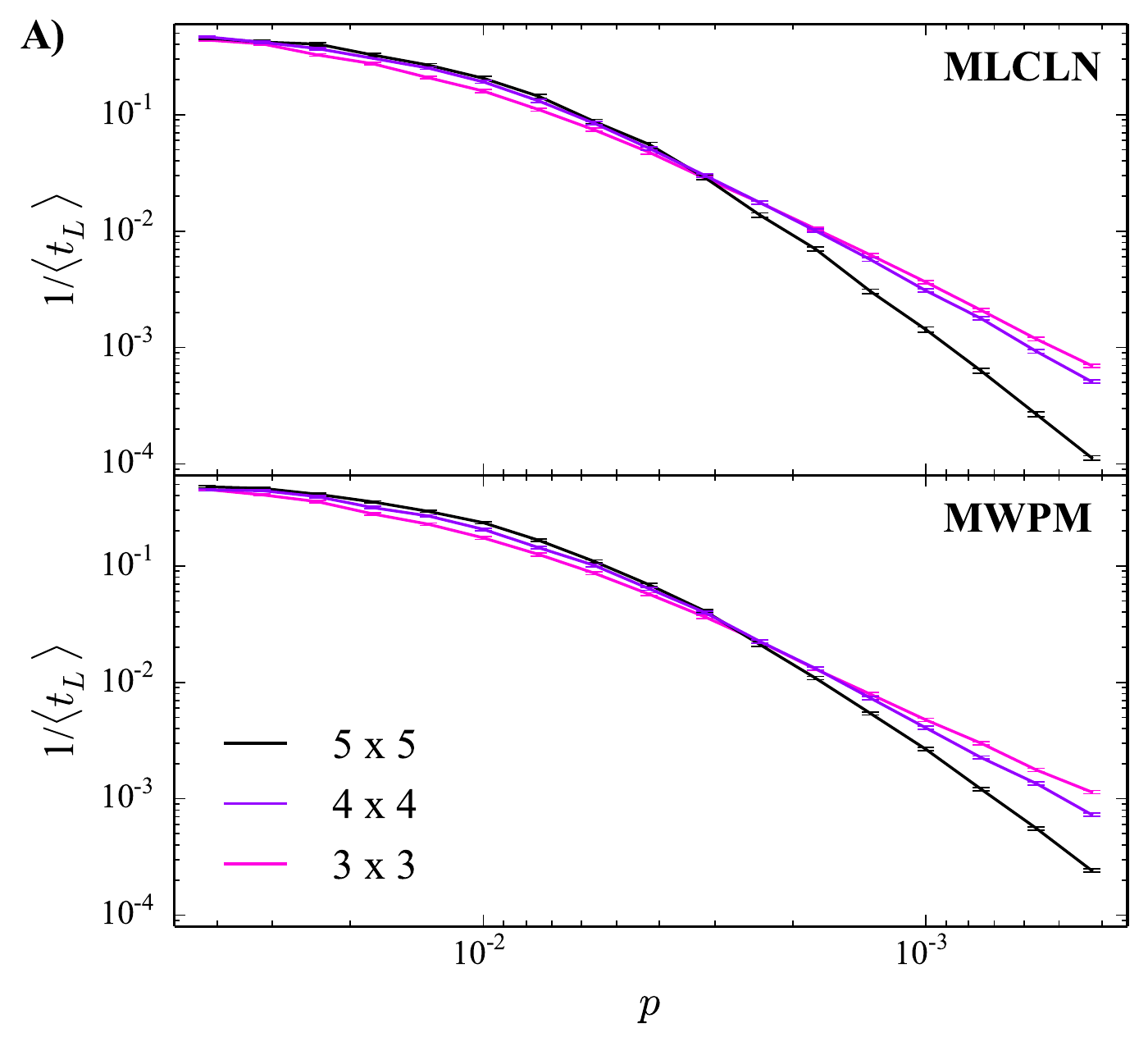}	
	\hspace*{-0.22cm}\includegraphics[width=1.04\columnwidth]{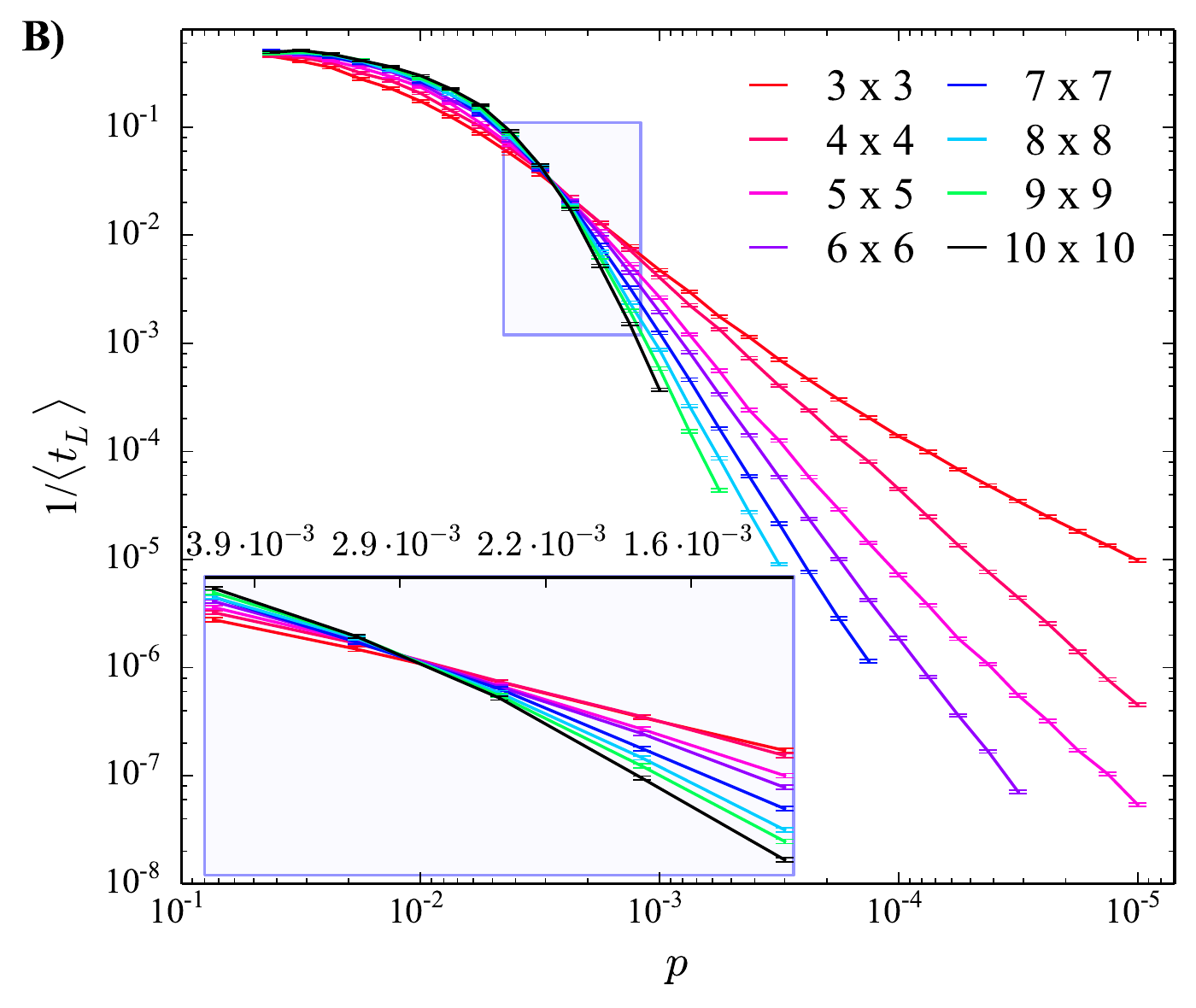}
	\vspace*{-0.7cm}\caption{Error threshold for independent bit and phase flip errors on each gate and qubit, using an 8-step syndrome measurement circuit. A) Comparison between MWPM and MLCLN. B) Extended plot for MWPM.}
	\label{IndependentXZthreshold}
	\vspace{-0.5cm}
\end{figure}	

Though the high value is cause for optimism regarding the implementation on existing hardware, one should be cautious. An 8-step measurement circuit is required if one cannot directly initialize and measure in the $X$-basis.
Additionally, within this noise model the probability for two qubits to have at least one error is only half as large when they are in a $CNOT$ gate compared to when they are idle;
	even though the probability of both qubits being faulty after a $CNOT$ gate is higher, such correlated errors are potentially easier to handle for the decoder, particularly for MLCLN.
	One might wonder whether the threshold remains distinctly higher for optimal decoding compared to MWPM under uncorrelated noise.
	
Since in a realistic setting, error propagation leads to correlations even if the underlying noise on each gate acts independently on all qubits, we proceed to consider a minimally correlated model, where correlations arise solely due to propagation. For both single- and two-qubit gates, each qubit experiences independent bit- and phase-flip noise, {\it i.e.}, the probability for $X$ and $Z$ errors is $p(1-p)$ and the probability for $Y$ errors is $p^2$. 
	We refer to the supplemental material for data on completely correlation-free phenomenological noise, where any advantage of MLCLN stems from entropic effects.
		
The error threshold for an 8-step measurement circuit is shown in Fig.~\ref{IndependentXZthreshold}A. The extensive computational resources required limit the simulation for MLCLN to a distance five surface code. Even when decoding only bit-flip errors, the next larger code size requires updating a vector of length $2^{35}$ ($\sim 140 GB$ of memory). Nevertheless, the three smallest code sizes give a decent indicator of the threshold. For a more accurate estimate we compare Fig.~\ref{IndependentXZthreshold}A with an extended plot containing larger system sizes with MWPM in Fig.~\ref{IndependentXZthreshold}B. There seems to be only a minor drift of the crossing point towards higher probabilities for larger code sizes. This brings us to an estimate of $p_c = 0.28\%$ for MWPM compared to $p_c^{opt} = 0.35\%$ for MLCLN. 
The longer syndrome extraction circuit is only partially responsible for the diminished threshold compared to the noise model in Fig.~\ref{correlated_noise}, for which the longer circuit still results in a visibly higher value \cite{SM} even when bearing in mind that the probability of a bit-flip error only amounts to $2/3$ of it.
	The threshold remains distinctly different for an optimal decoding, despite the MWPM ideal noise type. Nonetheless, the somewhat smaller difference between MWPM and MLCLN may indicate that strong correlations impact the performance of MWPM. 


\begin{figure}
	\hspace*{-0.22cm}\includegraphics[width=1.04\columnwidth]{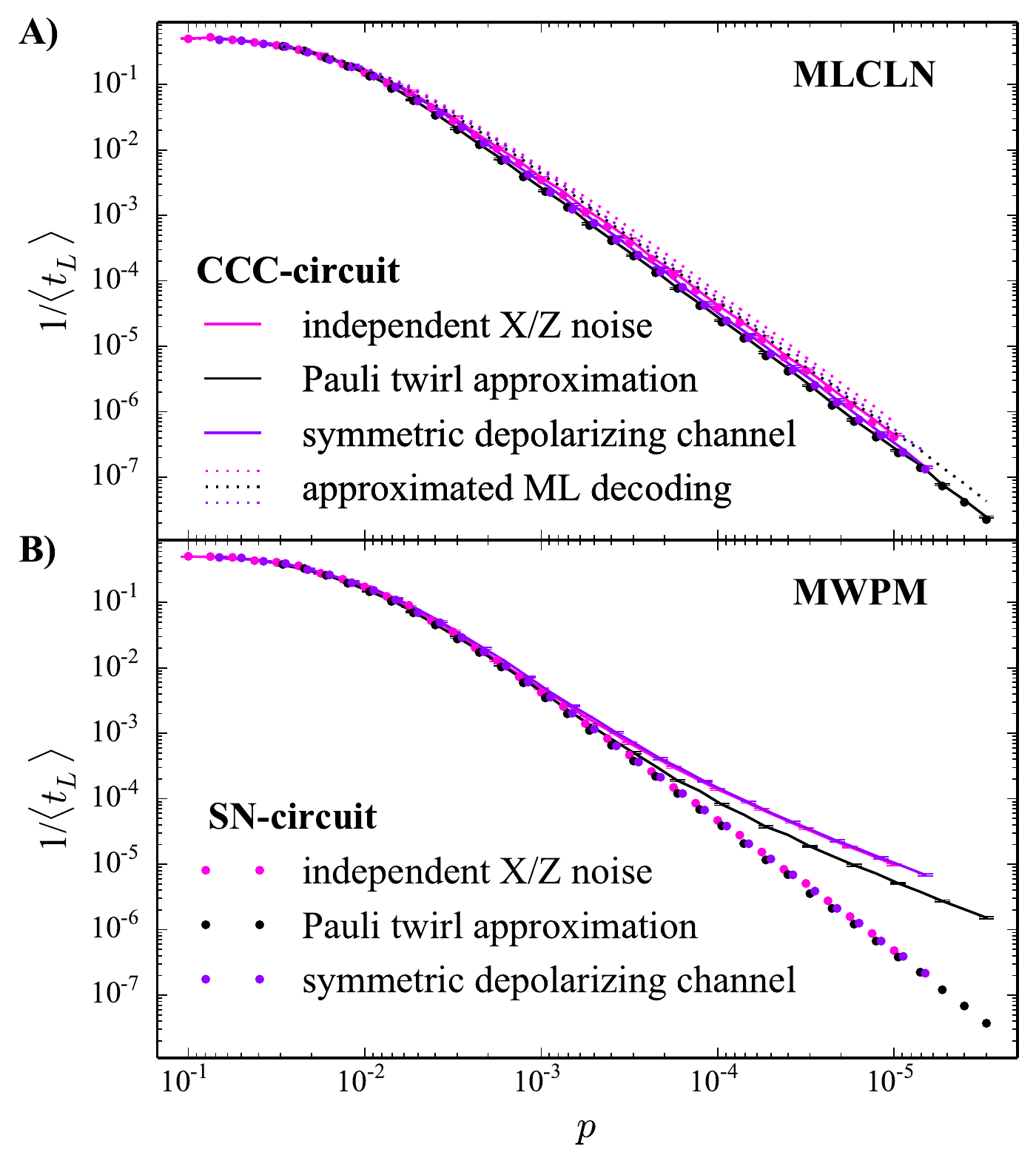}	
	\vspace*{-0.7cm}\caption{Analysis of the impact of correlations for a $3\times3$ surface code. To relate various noise models the x-axis denotes the (average) probability of an $X$ or $Y$ error on each qubit during one step of the measurement circuit. The legend belongs to both panels and has been split for better readability; solid lines in both panels indicate a CCC syndrome measurement circuit, and dots an SN-circuit.}
	\label{noise_models}
	\vspace{-0.5cm}
\end{figure}

{\it Impact of Error Correlations---}
To test that hypothesis, we analyze three types of correlations, all of which can only be partially captured by MWPM or not at all. The first are correlations between different errors on the same qubit. To that end we look at different noise channels that act independently on each qubit. The second are correlated errors between data qubits, which we analyze by comparing a syndrome measurement circuit that minimizes error propagation to one that spreads errors. The third are correlations between data and ancilla qubits.  
	
All simulations are shown in Fig.~\ref{noise_models}.
As long as noise acts independently on each qubit and error propagation is limited (SN-circuit), the results with MWPM match fairly well for all noise types, see Fig.~\ref{noise_models}B.
	Comparing the results for different noise models with MLCLN (Fig.~\ref{noise_models}A) suggests that a slight but minor improvement is possible using information about $Y$-correlations.
	Additionally, MLCLN exhibits no detectable performance drop even if the actual failure rates vary within $\pm 10\%$ of the ones assumed for decoding; using overly detailed knowledge about noise type and strength therefore does not seem to be essential 
		to optimize performance, as long as there are no correlated errors involving several qubits.


We proceed to noise where errors on different qubits are no longer independent.
To appraise its influence beyond what we noted when comparing Figs.~\ref{correlated_noise} and \ref{IndependentXZthreshold}, we opt to test different syndrome measurement circuits, comparing a CCC-circuit in which $CNOT$-gates are applied in a counter-clockwise (clockwise) order for X-stabilizers (Z-stabilizers) with an SN-circuit as proposed in Ref.~\onlinecite{Tomita2014}. 
	In contrast to an SN-circuit, a CCC-circuit is suboptimal in the sense that it does not minimize error propagation; higher order errors are more likely. 
	Its effect on the asymptotic behavior of MWPM, observed in Fig.~\ref{noise_models}B, suggests that strongly correlated noise or non-transversal logical gates may severely impact MWPM for low error rates. While this behavior is most pronounced for small code sizes, it persists for larger distances.
	An SN-circuit for a distance five and six code, for example, leads to a slope of -2.76 and -3.12 respectively for the exponential decline of the logical error rate, compared to values of -2.36 and -2.97 using a CCC-circuit.
	The performance drop of MWPM for certain multi-qubit errors is ultimately rooted in its lacking mechanism to match larger groups of defects. 
	As the results for MLCLN demonstrate, it can be avoided if we are able to use full knowledge about occurring correlations, see Fig.~\ref{noise_models}A.
	
The somewhat better values in the Pauli twirl curve with MWPM for a CCC-circuit can be explained by the chosen gate durations. It approximates amplitude damping noise by a Pauli channel \cite{Gosh2012}, 
	for which we chose gate durations of $20\mu s$ for single-qubit gates, $30 \mu s$ for two-qubit gates and $300 \mu s$ for measurements. 
The noise generation is favorably biased in the sense that most errors occur at the very end of the circuit. Even though the measured syndrome reflects them only after a one round delay, most defect pairs can be matched within the same cycle. 
This leads to the question whether taking correlations between faulty syndrome bits and data qubit excitations into account is vital for an good decoder performance. 

To test the importance of 
	relating data qubit errors to syndrome imperfections, we examine a maximum likelihood decoder based on a phenomenological noise model \cite{ColorCodesML}, which neglects exactly this kind of correlations.
	The error patterns at the end of the syndrome extraction circuit follow a certain global distribution of errors on the entire system of data plus ancilla qubits. Within a phenomenological model, this distribution is marginalized over data and ancilla qubits, respectively.  
	A decoder update then consists of applying a modified transition matrix that describes the spread of errors independent of the measured syndrome, followed by multiplication with a suitable error probability for faulty syndrome bits \cite{SM}.	
Despite this approximation, the asymptotic behavior remains unaffected by the CCC-circuit, see the dotted lines in Fig~\ref{noise_models}A. We deduce that the phenomenological approach of MWPM poses no fundamental limitation. 
The suspicion that mainly strong correlations between errors on multiple data qubits are difficult to process with MWPM seems thus warranted.  


{\it Conclusion---} 
	We have presented the first optimal decoder for circuit-level noise. We determined a maximal error threshold of 1.8\% for the standard noise model in Ref.~\onlinecite{WangFowler}, and 0.35\% for an independent bit- and phase-flip channel. 
	We find non-negligible performance improvements even under almost correlation-free noise, suggesting that the true threshold value is systematically underestimated by previous publications based on MWPM. 
	We have analyzed the impact of correlations on decoder performance for both optimal and MWPM decoding. We find that MWPM is close to optimal for all considered types of independent noise, but can be crucially improved when multi-qubit errors are as likely as single-qubit errors. Decoding based on a phenomenological model relies on the model matching the circuit noise reasonably well \cite{SM}, but overall does not seem to significantly impair performance.
	Our results indicate that correlations between data qubit errors have the most significant impact.
	It remains an open question whether such correlations can be sufficiently incorporated into renormalization group decoding schemes \cite{RG1,RG2,RG3} that outperform the roughly quadratic runtime scaling of MWPM with system size \cite{Fowler2012, fowler2012timing}.
	
	We thank the Quantum Architectures and Computation group members for helpful discussions.


\newpage
\bibliography{biblio}{}

\end{document}